\documentclass[aps,prl,twocolumn,secnumarabic,amsmath,amssymb,superscriptaddress,longbibliography]{revtex4-1}
\usepackage[colorlinks,bookmarks=true,citecolor=blue,linkcolor=red,urlcolor=blue,citecolor=blue]{hyperref}

\usepackage{graphicx}
\usepackage{xcolor}
\usepackage{subfigure}
\usepackage{appendix}
\usepackage{amssymb,amsmath}
\usepackage{color}
\begin{document}

\title{Umklapp scattering in the one-dimensional Hubbard model}

\author{Tong Liu}
\affiliation{Institute of Physics, Chinese Academy of Sciences, Beijing 100190, China.}

\affiliation{School of Physical Sciences, University of Chinese Academy of Sciences, Beijing 100049, China.}

\author{Kang Wang}
\affiliation{Institute of Physics, Chinese Academy of Sciences, Beijing 100190, China.}
\affiliation{School of Physical Sciences, University of Chinese Academy of Sciences, Beijing 100049, China.}

\author{Runze Chi}
\affiliation{Institute of Physics, Chinese Academy of Sciences, Beijing 100190, China.}
\affiliation{School of Physical Sciences, University of Chinese Academy of Sciences, Beijing 100049, China.}

\author{Yang Liu}
\affiliation{Institute of Physics, Chinese Academy of Sciences, Beijing 100190, China.}
\affiliation{School of Physical Sciences, University of Chinese Academy of Sciences, Beijing 100049, China.}

\author{Haijun Liao}\email{navyphysics@iphy.ac.cn}
\affiliation{Institute of Physics, Chinese Academy of Sciences, Beijing 100190, China.}
\affiliation{Songshan Lake Materials Laboratory, Dongguan, Guangdong 523808, China.}

\author{T. Xiang}\email{txiang@iphy.ac.cn}
\affiliation{Institute of Physics, Chinese Academy of Sciences, Beijing 100190, China.}
 \affiliation{Beijing Academy of Quantum Information Sciences, Beijing, 100190, China.}

\affiliation{School of Physical Sciences, University of Chinese Academy of Sciences, Beijing 100049, China.}

\begin{abstract}
  The Mott metal-insulator transition is a typical strong correlation effect triggered by the Umklapp scattering. However, in a physical system, the Umklapp scattering coexists with the normal scattering, including both forward and backward scattering, which conserves the total momentum of scattered electrons. Therefore, it is not easy to quantify the contribution of the Umklapp scattering in a Mott metal-insulator transition. To resolve this difficulty, we propose to explore these scattering processes separately. We study the contribution of each scattering process in the one-dimensional Hubbard model using the momentum-space density-matrix renormalization group (kDMRG) and bosonization methods. Our kDMRG calculation confirms that the Mott charge gap results from the Umklapp scattering, but the normal scattering processes strongly renormalize its value. Furthermore, we present a scaling analysis of the Mott charge gap in the bosonization theory and show that the interplay between the Umklapp and forward scattering dictates the charge dynamics in the half-filled Hubbard model.
\end{abstract}

\maketitle

\textbf{Introduction:} Mott insulator has gained intensive attention from condensed matter physicists\cite{Boer_1937,mott1990insulator, gebhard1997mott,mottness,1993spectral_transfer,luttingersurface,charge2e,mottRMP}, not only because Mott insulator serves as a platform for studying strong correlation effects, but also by doping Mott insulator, we can get many novel phases, like high-$T_c$ superconductor, pseudogap, non-fermi liquid, charge density wave, etc\cite{Markiewicz2008,pseudogap,Science2000,nonFermi,charge_excitation2002,dopingmott}.   This interaction-driven effect  could be understood qualitatively using the Hubbard model at half-filling~\cite{hubbard1963electron,Hartree-Fock}. The Hubbard system is metallic in the weak coupling limit, where the onsite Coulomb interaction is small compared to the kinetic energy. However, in the strong coupling limit, the onsite Hubbard interaction dictates the conducting behaviors. It tends to localize electrons by raising the energy of double occupation on a single lattice site, which opens a charge excitation gap at half-filling. Therefore, the half-filled Hubbard system undergoes a Mott metal-insulator transition from the weak to strong coupling limit.

 The Hubbard interaction can be decomposed into three terms according to their scattering processes: forward, backward, and Umklapp scattering. Both the forward and backward scattering processes conserve the total momentum of electrons. They correlate electrons in the ferromagnetic and antiferromagnetic channels, respectively. However, the Umklapp process conserves the total momentum of scattered electrons up to a reciprocal-lattice vector. This process is greatly enhanced around the half-filling and is the driving force that is responsible for the formation of the Mott insulating gap. In the band theory, a half-filled band is a metal rather than an insulator. However, the Umklapp scattering bounces two electrons in the vicinity of one side of the Fermi surface to the opposite side in one dimension, leading to a charge excitation gap with divergent charge compressibility at half-filling~\cite{giamarchi1991umklapp, mori1994towards, schulz1990correlation, emery1990strong}. The Umklapp process might be responsible for the pseudogap phenomenon observed in high-T$_c$ copper oxides~\cite{rice2017umklapp, robinson2019anomalies}. It can also induce a topologically nontrivial edge state~\cite{klinovaja2013topological}.

 A thorough investigation of the Umklapp scattering is essential to a qualitative understanding of the Mott physics \cite{HonerKamp2001,Wu2017,Phenomenological2006,Phenomenological2012,twochainhubbard,2dhubbardRG}. However, it is difficult to investigate the Umklapp process because it coexists with the forward and backward processes in real materials. Their interplay makes it hard to unveil the secret of the Umklapp scattering. Nevertheless, in theoretical studies, we can separate these scattering processes and consider the contribution of each process independently, allowing us to quantitatively investigate the effect of the Umklapp scattering on the Mott insulating transition and how it interferes with other processes.

In this work, we present a comparative study of the Hubbard model and a modified Hubbard model, which contains only the Umklapp scattering term, namely ignoring the forward and backward scattering terms, in the on-site Coulomb interactions. We call this modified Hubbard model the Umklapp model. The one-dimensional Hubbard model is soluble by the Bethe Ansatz~\cite{lieb1994absence,lieb1968absence}.  It can also be accurately probed by  the real-space density-matrix renormalization group (DMRG)~\cite{daul1996dmrg,arita1998density} and quantum Monte Carlo~\cite{sandvik1993quantum}. However, it is much more challenging to solve the Umklapp model. First, the Bethe Ansatz does not work for this model. Furthermore, there are technical barriers in carrying out real-space DMRG and quantum Monte Carlo simulations for this model because the Umklapp scattering potential is long-ranged and suffers from the minus-sign problem even at half-filling.

We propose to use the momentum-space DMRG (kDMRG)~\cite{xiang1996density}, combined with a scaling analysis of the coupling constants in the framework of bosonization~\cite{emery1979highly}, to resolve the above difficulties. kDMRG is an effective method for exploring this problem because the Umklapp scattering potential takes a relatively simple representation in momentum space. The scaling analysis, on the other hand, allows us to gain a more clear picture on how the interplay between different scattering processes affects the Mott insulating behavior~\cite{emery1976solution,giamarchi1991umklapp}.

\textbf{Model:} The one-dimensional Hubbard model is described by the Hamiltonian:
\begin{eqnarray}
    H&=&-t\sum_{j\sigma } \left( c^{\dagger}_{j\sigma} c_{j+1,\sigma} +h.c. \right)+ U\sum_j n_{j\uparrow}n_{j\downarrow} ,
    \label{Eq:Hubbard}
\end{eqnarray}
where $c_{j\sigma}$ is the annihilation operator of electron at site i, and $n_{j\sigma} = c_{j\sigma}^\dagger c_{j\sigma}$. In momentum space, it becomes
\begin{eqnarray}
    H&=& -2t\sum_{k \sigma } \cos k c^\dagger_{k\sigma}c_{k\sigma} + H_n + H_u     \\
    H_n&=&\frac{U_1}{L}\sum_{k_1k_2k_3k_4} c^{\dagger}_{k_1\uparrow} c_{k_2\uparrow} c^{\dagger}_{k_3\downarrow} c_{k_4\downarrow} \delta_{k_1+k_3,k_2+k_4} ,\\
    H_u&= & \frac{U_2}{L}\sum_{k_1k_2k_3k_4} c^{\dagger}_{k_1\uparrow} c_{k_2\uparrow} c^{\dagger}_{k_3\downarrow} c_{k_4\downarrow}\delta_{k_1+k_3,k_2+k_4\pm 2\pi} .
\end{eqnarray}
 Here we separate the Coulomb interaction terms into two parts according to the scattering processes. $H_n$ is the Hamiltonian of normal scattering, including both forward and backward scattering, which preserves the total momentum. $H_u$, on the other hand, is the Hamiltonian of the Umklapp scattering, which preserves the total momentum up to a reciprocal lattice vector. To distinguish these terms explicitly, we assume $H_n$ and $H_u$ to have different coupling constants, $U_1$ and $U_2$.

 The Hubbard model (\ref{Eq:Hubbard}) corresponds to the case $U_1=U_2=U$. We can screen the normal scattering process by setting $U_1=0$. In that case, $H$ is just the Hamiltonian of the Umklapp model. Similarly, we can switch off the Umklapp scattering by setting $U_2=0$ and refer to the resulting Hamiltonian as the non-Umklapp model. The Hubbard interaction is local in real space. However, for the above generalized Hubbard model, the interaction becomes highly non-local when transformed back from the momentum-space representation to real space in the case $U_1\not= U_2$.

\begin{figure}[t]
\centering
\includegraphics[width=0.43\textwidth]{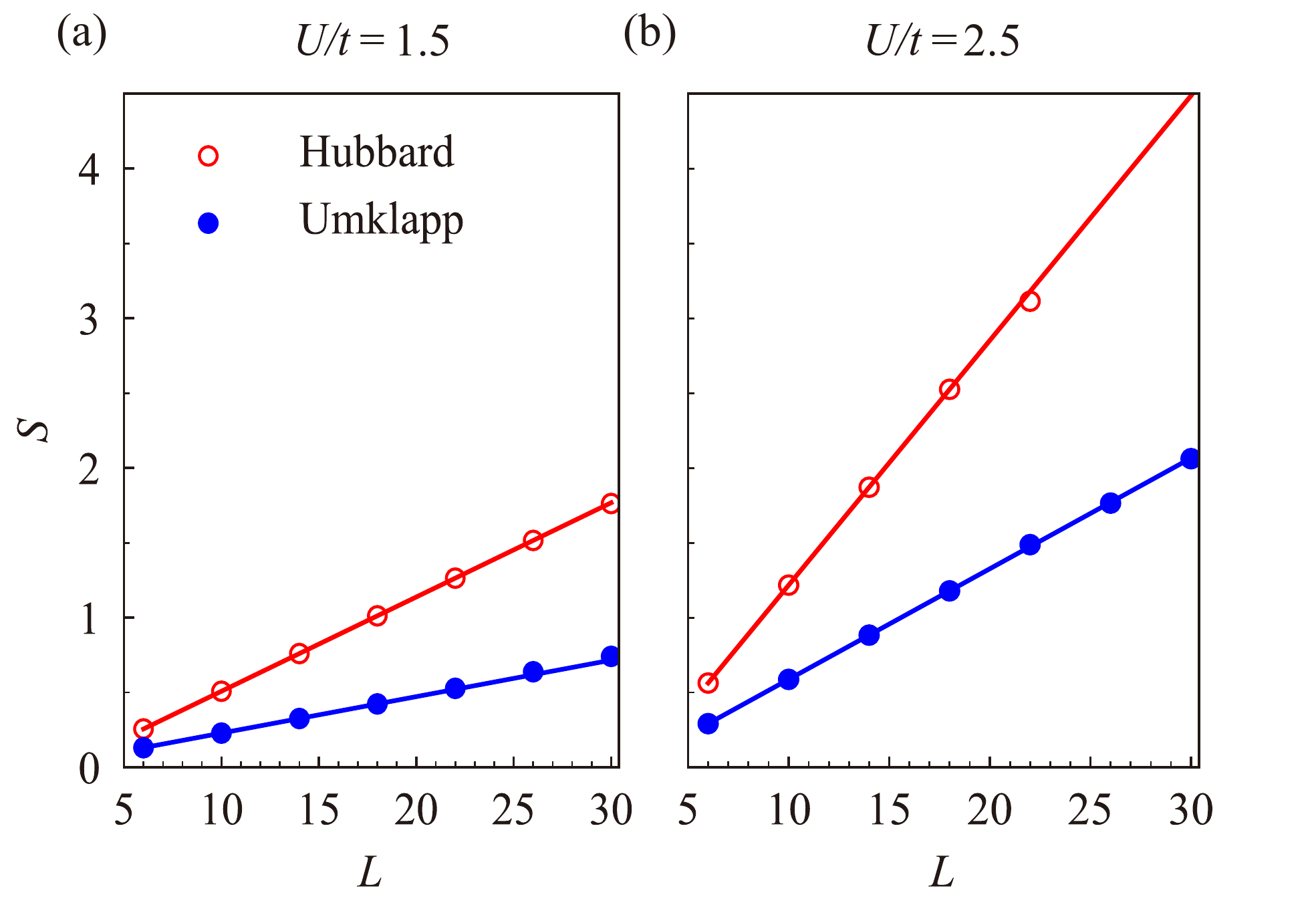}
\caption{
Comparison of the size dependence of the maximal entanglement entropy $S$ calculated by kDMRG for the ground state of the Hubbard model with that of the Umklapp model ($U_2=U$). }

\label{Fig:Entropy}
\end{figure}

\textbf{Results of kDMRG:}
In kDMRG, we set a momentum-spin point as a lattice site. A crucial step in the kDMRG calculation is to find an optimized path to order these momentum-spin points in a one-dimensional chain. As discussed in the supplementary materials, we determine this optimized path by minimizing a mutual-information distance of all momentum-spin points. To avoid a numerical instability induced by the degeneracy in the kinetic energy, we only consider systems of size $L=4n+2$ ($n$ an integer) with periodic boundary conditions.

In momentum space, it is known that the entanglement entropy $S$ of the ground state scales linearly with the system size $L$ for the Hubbard model~\cite{ehlers2015entanglement}. Hence the ground state of the Hubbard model satisfies an entanglement volume law in momentum space. Our kDMRG calculation confirms this volume-law behavior of the entanglement entropy for the Hubbard model.

Figure~\ref{Fig:Entropy} shows the entanglement entropy $S$ of the ground state obtained from the kDMRG calculations for the Umklapp and Hubbard models. For the Umklapp model, the entanglement entropy $S$ also scales linearly with the system size. However, the entanglement entropy of the Umklapp model is much lower than that of the Hubbard model. For the two cases shown in Fig.~\ref{Fig:Entropy}, the entanglement entropy of the Umklapp model is about half of the Hubbard model. It implies that one can reliably study much larger lattice systems for the Umklapp model than for the Hubbard model using kDMRG by keeping the same number of basis states.

The key parameter characterizing an Mott insulating phase is the charge excitation gap, $\Delta_c$, defined by the energy increase in adding and removing a pair of spin-singlet electrons from the half-filled system:
\begin{equation}
    \Delta_c(L) =  \frac{1}{4}[E_g(L+2)+ E_g(L-2) - 2E_g(L)]
\end{equation}
where $L$ is the lattice size and $E_g(N_e)$ is the ground state energy of the system with the electron number $N_e$. $N_e$ equals $L$ at half-filling.

\begin{figure}[t]
\centering
 \includegraphics[width=0.4\textwidth]{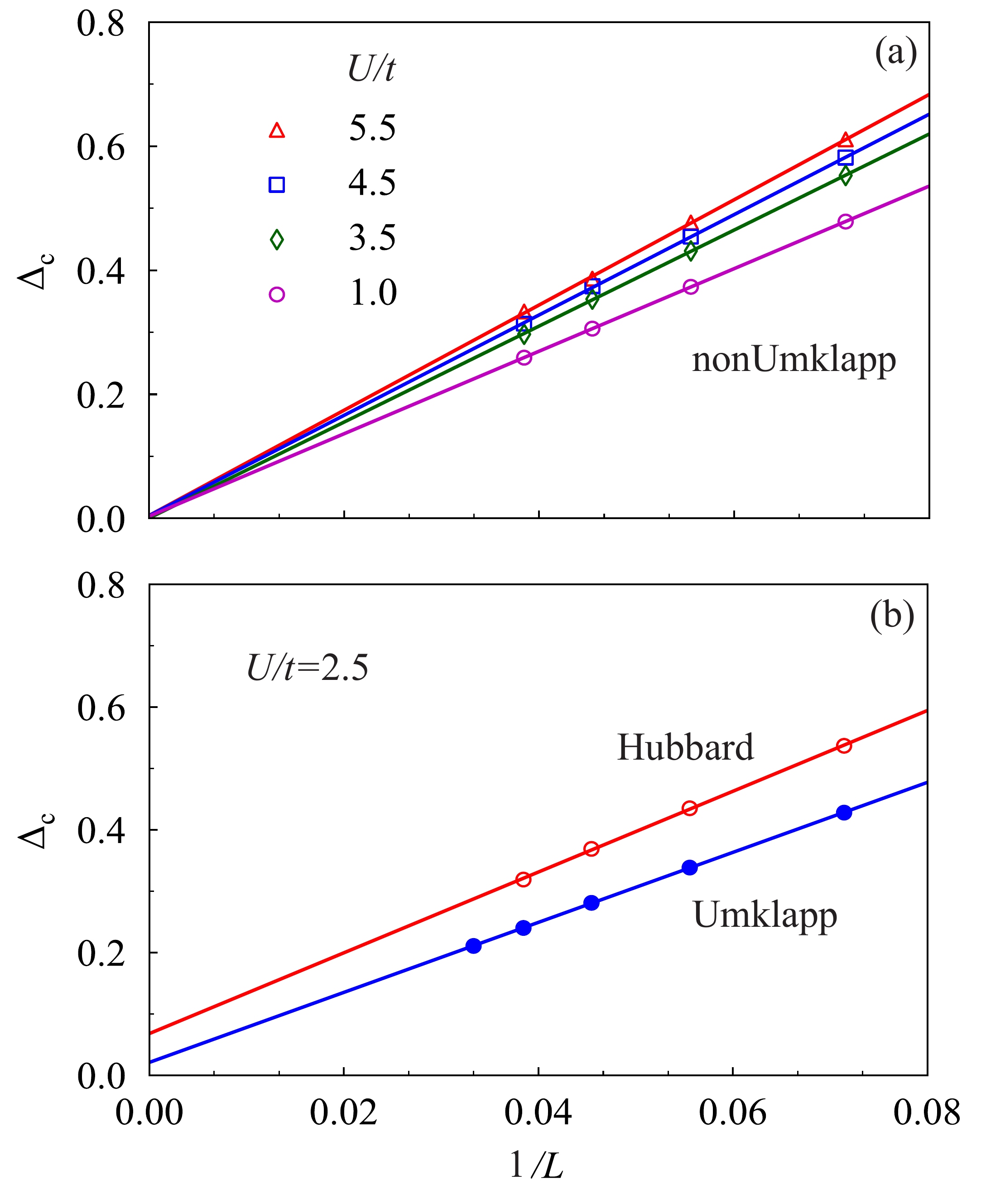}
\caption{
 Charge excitation gap $\Delta_c$ as a function of $1/L$ for (a) the non-Umklapp model ($U_1=U$ and $U_2=0$) and (b) the Hubbard and Umklapp ($U_1=0$ and $U_2=U$) models obtained with kDMRG. The lines are linear fits to the kDMRG data.
}
\label{Fig:Umklapp}
\end{figure}

Figure~\ref{Fig:Umklapp}(a) shows the kDMRG result of $\Delta_c$ as a function of the inverse lattice length $1/L$ in the absence of the Umklapp scattering ($U_2=0$). In this case, $\Delta_c$ scales linearly with $1/L$ within numerical errors. By linear extrapolation, we find that the gap excitation gap vanishes in the thermodynamic limit. Thus the system remains gapless no matter how strong the normal scattering interaction $U_1$ is.

 However, in the presence of the Umklapp scattering, the charge excitation spectrum is gapped. Figure~\ref{Fig:Umklapp}(b) compared the size dependence of the charge gap $\Delta_c$ for the Hubbard and Umklapp models at half-filling. Again, $\Delta_c$ scales linearly with $1/L$ within numerical errors, but the extrapolated gap value in the thermodynamic limit $L\rightarrow \infty$ is finite. Moreover, the Umklapp scattering also changes the momentum distribution of electrons when two more electrons are added to the half-filled system. As shown in Fig.~\ref{Fig:nk} in the supplementary material, the momentum distribution function is mirror symmetric about the $k=0$ point in the Hubbard model, and the total momentum of the ground state is zero. For the Umklapp model, however, the two added electrons tend to have the same momentum, which breaks the mirror symmetry in the momentum distribution function
 
 The above discussion confirms that the Mott insulating gap arises from the Umklapp scattering rather than the normal scattering processes. However, by comparing the gap value of the Umklapp model with that of the Hubbard model, we find that the normal scattering processes can significantly enhance the value of the Mott gap once it is open.

 \textbf{Scaling analysis:} To understand the physics underlying the enhancement of the Mott charge gap by the normal scattering, we perform a scaling analysis for the generalized Hubbard model in the bosonization theory. In the long-wavelength limit, the charge and spin excitation spectra in the Hubbard model are separated and effectively described by two boson fields. Following the standard bosonlization scheme, it is straightforward to show that the following two Hamiltonians govern the charge and spin dynamical properties of the Hubbard model:
\begin{eqnarray}
 H_c&= & \frac{v_F a}{2} \int d x \Big[ \Pi_c^2+ \left(1+\frac{U_1}{\pi v_F} \right)(\partial_x \phi_c)^2  \nonumber \\
 && \hspace{1.8cm} +  \frac{U_2 \cos\sqrt{8\pi} \phi_c}{v_F \pi^2 \alpha^2} \Big] , \label{Eq:Hc}\\
 H_s&= &\frac{v_F a}{2}\int d x \Big[ \Pi_s^2+ \left( 1-\frac{U_1}{\pi v_F} \right) (\partial_x \phi_s)^2 \nonumber \\
 && \hspace{1.8cm} +\frac{U_2 \cos\sqrt{8\pi} \phi_s}{v_F \pi^2 \alpha^2} \Big],
 \label{Eq:Hs}
\end{eqnarray}
where $\phi_c$ and $\phi_s$ are the boson fields in the charge and spin channels, respectively. $\Pi_\sigma$ ($\sigma = c$, $s$) is the conjugate field of $\psi_\sigma$. $\alpha$ is the inverse of the momentum cutoff.

In $H_c$, the $U_1$ term results from the forward scattering terms in $H_n$. However, the $U_1$ term in $H_s$ is the contribution of the backward scattering. Thus the backward scattering affects the spin dynamics but not the charge dynamics. These $U_1$ terms renormalize the charge and spin velocities to
\begin{equation}
 v_c=v_F \left( 1+\frac{U_1 a}{2 \pi v_F} \right), \quad v_s=v_F \left( 1-\frac{U_1 a}{2 \pi v_F} \right).
\end{equation}

The scaling dimensions of the cosine terms in charge and spin channel are now given by
\begin{equation}
d_c=\frac{2}{ \sqrt{ 1+\frac{U_1}{\pi v_F} } }, \quad
d_s=\frac{2}{ \sqrt{ 1-\frac{U_1}{\pi v_F} } }.
\end{equation}
For a positive $U_1$, the scaling dimension of the charge field $d_c <2$. In this case, the Umklapp term is relevant. It opens a gap in the charge excitation spectrum. Qualitatively speaking, the smaller $d_c$ (or larger $U_1$), the larger the charge gap $\Delta_c$. Thus the forward scattering can enhance the charge gap. On the contrary, the scaling dimension of the spin field $d_s > 2$ and the corresponding Umklapp scattering term is irrelevant. Consequently, the spin excitation remains gapless.

\begin{figure}[t]
\centering
\includegraphics[width=0.4\textwidth]{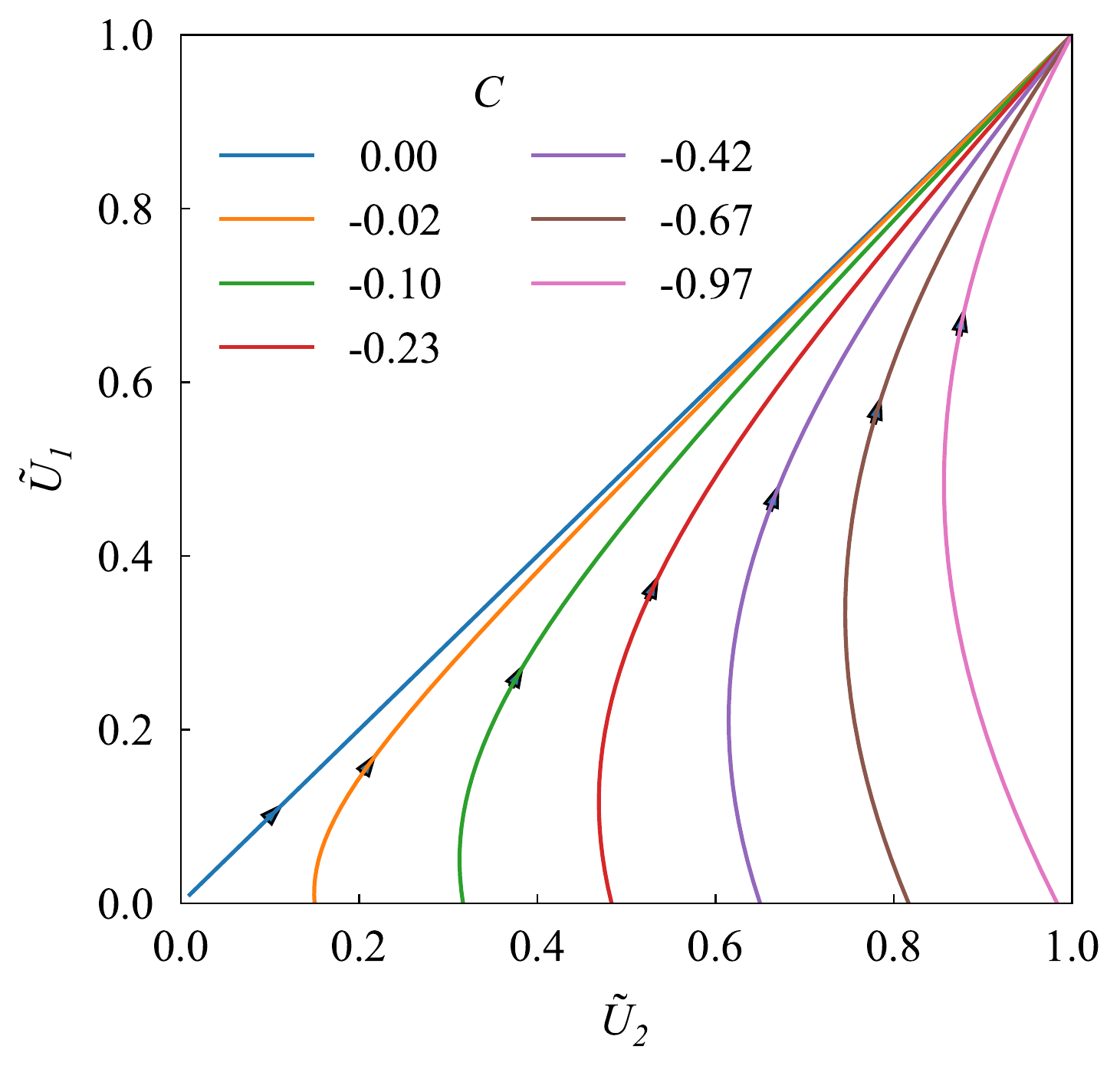}
\caption{RG flow of the coupling constants $\tilde U_1$ and $\tilde U_2$. }
\label{Fig:RGFlow}
\end{figure}

 To quantitatively understand how the charge gap varies with the coupling constants, let us consider the scaling behavior of these parameters under the renormalization-group (RG) transformation. Larkin and Sak derived the RG equations of $U_1$ and $U_2$ to the third order of perturbation in the charge channel~\cite{larkin1977boundary,guo2003charge}. They obtained the following equations that govern the RG flow of $U_1$ and $U_2$ under the change of the scaling parameter $l$ in the momentum cutoff $\alpha \rightarrow \alpha^{\prime} = \alpha \exp ( d l)$,
\begin{eqnarray}
 \frac{d\tilde{U}_{1}(l)}{dl}&=& 2\tilde{U}_{2}^2(l) \left[ 1-\tilde{U}_{1}(l) \right] ,
 \label{RG_eq1} \\
 \frac{d\tilde{U}_{2}(l)}{d l} &=& 2\tilde{U}_{1}(l)\tilde{U}_{2}(l) -\tilde{U}_{1}^2(l)\tilde{U}_{2}(l)-\tilde{U}_{2}^3(l),
 \label{RG_eq2}
\end{eqnarray}
where $\tilde{U}_{i}= U_i/(2 \pi v_c)$ ($i=1$, $2$).

From the above equations, it is straightforward to show that the variable
\begin{equation}
 C=\frac{ \tilde{U}^2_1(l)-\tilde{U}^2_2(l) }{ 1-\tilde{U}_1(l) }
\end{equation}
is scaling invariant and the RG equation governing $\tilde U_1$ is
\begin{equation}
    \frac{d \tilde{U}_1(l)}{d l} = 2 \left[ \tilde{U}_1^2(l)+C\tilde{U}_1(l)-C \right] \left[ 1-\tilde{U}_1(l) \right] .
\label{Eq:running}
\end{equation}
 By solving this equation, we can find how the coupling constants $\tilde U_1$ and $\tilde U_2$ flow with $l$. The result, depicted in Fig.~\ref{Fig:RGFlow}, shows that a strong-coupling fixing point exists at $\tilde U_1 = \tilde U_2 =1 $  in the large $l$ limit: $U_1$ and $U_2$ always flow to this fixing point independent of their initial values.

%In a gapped system, the correlation length of the wave function in the charge channel is inversely proportional to charge gap. With the coupling constants flowing to the fixed point, the correlation length gets renormalized as well. In this way, if we can determine the charge gap $\Delta_c(U_{1s},U_{2s})$ for the system at specific $U_{1s},U_{2s}$ in the above phase diagram, then the charge gap $\Delta_c(U_{1},U_{2})$ on the whole RG trajectory can be determined.

  In a gapped system, the correlation length is inversely proportional to the charge gap and upper bound by the charge gap. Consequently, $l$ is also constrained by the charge gap $ l < l_s = \ln [ \beta v_c k_F / \Delta_c (U_1, U_2)] $, where $l_s$ is the increment of the coupling constants along the trajectory from $(U_1, U_2)$ to $(U_{1s}, U_{2s})$. $\beta v_c k_F$ is the characteristic energy scale of the system with $\beta$ a coefficient that depends on $(U_{1s}, U_{2s})$. $k_F = \pi /(2a)$ is the Fermi vector. From Eq.~(\ref{Eq:running}), we have
\begin{equation}
    \int_0^{l_s} dl=\int^{\tilde{U}_{1s}}_{\tilde{U}_1} \frac{du}{2(u^{2}+Cu-C)(1-u)} .
    \label{Eq:run2}
\end{equation}
Both integrals in the above equation are constrained along the trajectory with  $C=C(\tilde{U}_{1s},\tilde{U}_{2s})$. Solving Eq.~(\ref{Eq:run2}), we obtain the expression of the charge gap
\begin{equation}
   \Delta_c(U_1,U_2)=\beta v_c k_F e^{ F(\tilde{U}_{1}) - F(\tilde{U}_{1s}) } ,
   \label{Eq:gap}
\end{equation}
where
\begin{eqnarray}
F(x)&=&-\frac{1}{2}\ln |1-x|+\frac{1}{4} \ln \left( x^2+C x-C \right) +
\nonumber
\\
    &&\left\{
\begin{array}{ll}
\displaystyle
  \frac{C+2}{2C_1}\tanh^{-1}\frac{C+2x}{C_1}, & |C+2|>2
\vspace{2mm}
\\
\displaystyle
  \frac{C+2}{2C_1}\tan^{-1}\frac{C+2x}{C_1},& |C+2|<2
\vspace{2mm}
\\
\displaystyle
  -\frac{1}{2x} , & |C+2|=2
\end{array}
\right.
\end{eqnarray}
and $C_1=\sqrt{| C^2 + 4C|}$.

\begin{figure}[t]
\centering
\includegraphics[width=0.4\textwidth]{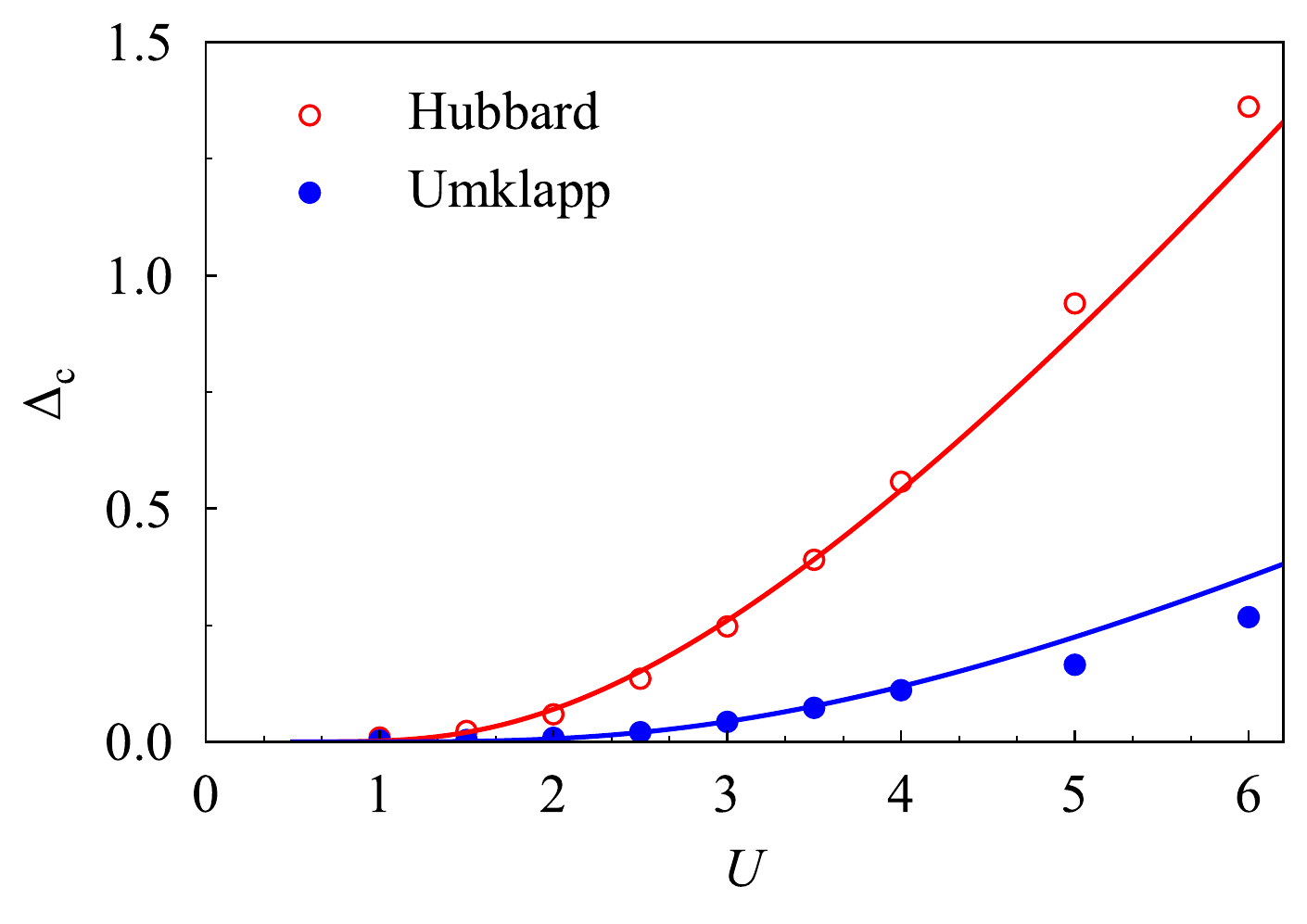}
 \caption{
 Charge gap $\Delta_c$ versus $U$ for the Hubbard and Umklapp ($U_2=U$) models obtained from the bosonization (solid curves) and kDMRG (open circles) calculations. The kDMRG data are obtained by extrapolating the kDMRG results at finite $L$ to the thermodynamic limit $L\rightarrow \infty$.
 }
\label{Fig:gap}
\end{figure}

The Umklapp model with different $U_2$ has different scaling invariant parameters $C$. We need to determine the value of $\beta$ at a given $U_{1s}$ for each RG trajectory. However, the difference in the value of $\beta$ between different trajectories at a given $\tilde{U}_{1s}$ is small in the limit $\tilde{U}_{1s}\rightarrow 1$ if $C\ll 1$.  Hence we can use the value of $\beta$ obtained by fitting the expression (\ref{Eq:gap}) on the trajectory $\tilde{U}_1=\tilde{U}_2$ ($C=0$) with the DMRG results shown in Fig.~\ref{Fig:gap} to determine the gap values of the Umklapp model. By taking $\tilde{U}_{1s} =0.99$, we find that $\beta = 11$. Using this parameter, we can estimate the gap values for the model with $\tilde U_1\not=\tilde U_2$  ($C\not= 0$).

  Figure~\ref{Fig:gap} compares the results of $\Delta_c$ obtained by kDMRG with those predicted by RG equations. The agreement between the results obtained with these two approaches is excellent, except in the strong coupling regime where the higher-order correction of perturbations to the RG equations should be considered. It confirms that the bosonization theory correctly catches up with the low-energy physics of the generalized Hubbard model, and the Mott insulating gap results from the Umklapp scattering of electrons around the Fermi level.
  
\textbf{Summary:} In summary, we have analyzed the role of different scattering processes on the Mott insulating transition by invoking both kDMRG and bosonization methods. From the kDMRG calculation, we obtain for the first time the charge excitation gap as a function of the coupling constant for the one-dimensional Umklapp model. By comparing the results of the Umklapp model with that of the Hubbard model, we show that the Mott insulating gap is triggered by the Umklapp scattering, as expected, and that the forward scattering can strongly renormalize the gap value. In one dimension, the backward scattering weakly affects the Mott insulating gap. However, in two dimensions, the backward scattering may induce a long-range antiferromagnetic order and strongly interfere with the Umklapp scattering. 

The interplay between different scattering processes enriches the physics of the Hubbard model. However, it also blurs the picture of the Mott metal-insulator transition, especially in two or higher dimensions, making it difficult to establish a quantum field theory description of the Mott insulator. In this work, we independently study the effect of each scattering process by screening some scattering processes in the one-dimensional Hubbard model. This strategy can be extended to two or higher dimensions. A project along this line is in progress. We hope it will allow us to capture the main physics governing the Mott metal-insulator transition and to find a general scheme for creating a quantum spin liquid by suppressing the antiferromagnetic order while keeping the Mott insulating gap open.

\textbf{Acknowledgement:} This work is supported by the National Key Research and Development Project of China (Grant No.~2017YFA0302901), 
the National Natural Science Foundation of China (Grants Nos.~11888101,~11874095, and~11974396), 
the Youth Innovation Promotion Association CAS (Grants No.~2021004), and the Strategic 
Priority Research Program of Chinese Academy of Sciences (Grant Nos.~XDB33010100 and~XDB33020300). 

\bibliographystyle{apsrev4-1}
\bibliography{Umklapp}

\begin{figure*}
\section{SUPPLEMENTARY MATERIAL}
\end{figure*}
\newpage
\label{sec:SM}
\textbf{Optimization of sites order:} The interacting potential of the Umklapp model becomes highly non-local in real space, and it is hard to investigate this model using the real-space DMRG. In momentum space, on the other hand, this complexity can be significantly reduced. More specifically, one can dramatically lower the computational cost by minimizing the number of operators whose matrix elements need to be evaluated and stored in kDMRG from $L^3$ to $6L$ using the regrouping technique first introduced in Ref.~\cite{xiang1996density} combined with the momentum conservation.

In momentum space, the lattice is a collection of all momentum-spin points ($k,\sigma$). Hence, a momentum-spin point now represents a lattice site. These momentum-spin points are ordered to form a one-dimensional lattice used for kDMRG calculations. In real space, the lattice sites have a natural order as the interactions are local. However, many ways exist to order the lattice sites in momentum space. Therefore, to optimize the kDMRG results, one needs first optimize the order of these momentum-spin points. We do this in two steps:

First, starting from a trial order of the momentum-spin points guessed based on physical intuition, we perform the standard kDMRG calculation to activate all the momentum-spin points. Then we sweep the lattice a couple of times. At each step, we swap two adjacent sites if the bipartite entanglement entropy between the left and right blocks separated by these two sites is lower than the case without swap.

Second, we calculate the mutual information between any two momentum-spin points, $M_{k \sigma,k^{\prime}\sigma^{\prime}} $, when the kDMRG sweep reaches the middle of the lattice. Then we rearrange all the lattice sites by minimizing the distance of two sites weighted by their mutual information:
\begin{equation}
    M_\mathrm{dist}=\sum_{k\sigma,k^{\prime}\sigma^{\prime}} \left\vert R(k \sigma, k^{\prime} \sigma^{\prime}) \right\vert^2 M_{k \sigma,k^{\prime}\sigma^{\prime}} ,
\end{equation}

where $R(k \sigma, k^{\prime} \sigma^{\prime})$ is the lattice distance between $(k \sigma)$ and $(k^\prime \sigma^\prime)$. This minimization can further optimize the order of the momentum-spin points, preventing the ground state from being trapped in a local minimum.

\begin{figure*}[htbp]
\centering
\includegraphics[width=1.0\textwidth]{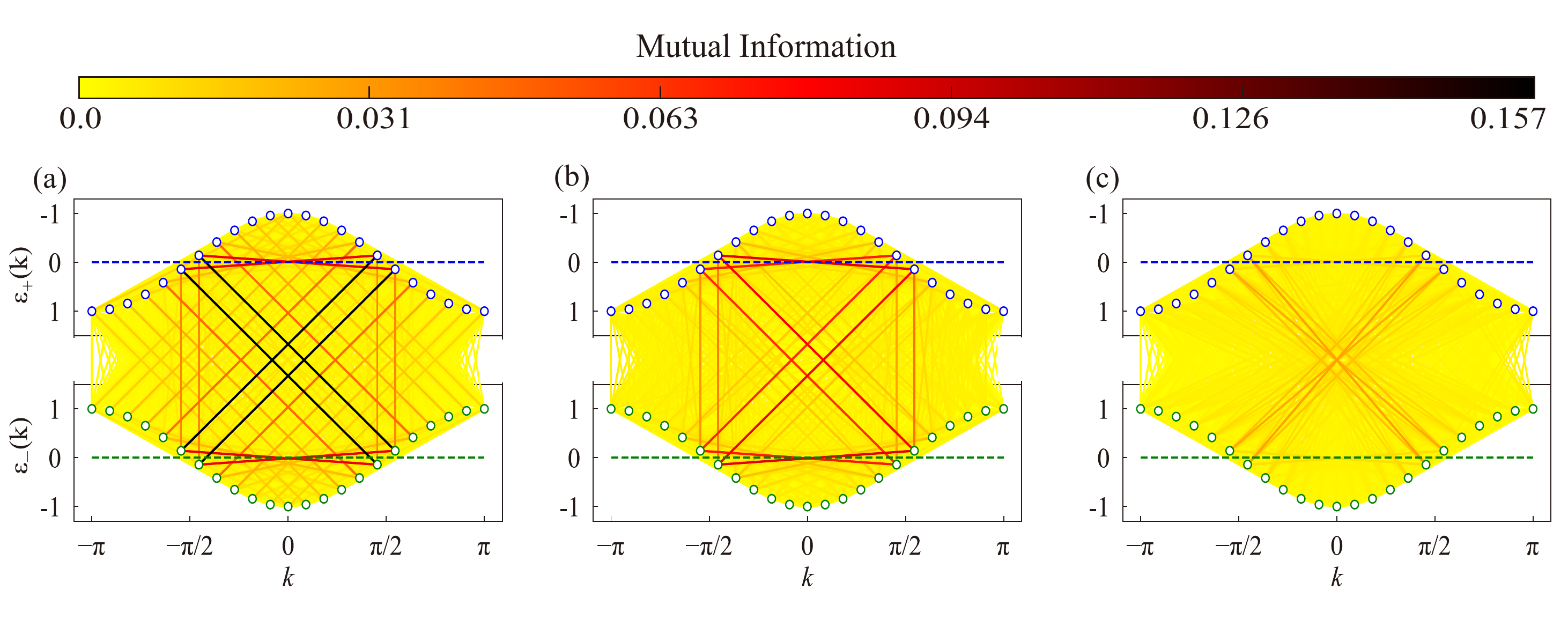}
\caption{
 Mutual information between two spin-momentum points in the ground states of the Hubbard, Umklapp ($U_2=U$), and $\pi$ non-Umklapp ($U_1=U$) models at half-filling.  $U=2.5t$ and $L=22$. $\varepsilon_k = -2t \cos k$ is the energy dispersion of electrons in the absence of interactions. The plot looks symmetric because the vertical axis of the energy dispersion for the up-spin electrons is reversed. The dashed lines represent the Fermi levels.
}
\label{Fig:MI}
\end{figure*}

\textbf{Mutual information structure:} 
Figure~\ref{Fig:MI} shows the intensity plot of the mutual information between two momentum-spin points in the optimized ground state for the Hubbard, Umklapp, and non-Umklapp models with $U=2.5$ at half-filling. The color scale of a line connecting two sites represents their mutual information. It is evident that electrons near the Fermi surface are most correlated in all three models, and the correlation between two electrons of different spins is stronger than that of the same spin.

For the Hubbard model, the checkerboard grid structure of the mutual information indicates that the two sites with a momentum separation $\pi$ have a more apparent correlation than other sites. For the Umklapp model, the correlation structure is similar. However, looking at the mutual information structure more carefully, we find that the correlation decreases gradually when the two momentum-spin points move away from the Fermi surface. Moreover, a stronger correlation is observed between two electrons with the same momentum.

The subtle difference in the mutual information between the Hubbard and Umklapp models also appears in the momentum distribution function of electrons. Figure~\ref{Fig:nk} shows the momentum distribution function $n_k$ for the two models with two more electrons added to the half-fill system. For the Hubbard model, two added electrons tend to have opposite momentum, and the momentum of the ground state is zero. Therefore, $n_k$ is mirror symmetric about the $k=0$ point. However, for the Umklapp model, the two added electrons tend to have the same momentum, and $n_k$ is non-symmetric with respect to the central reflection point $k=0$.

The correlation structure of the non-Umklapp model is distinctive from the former two models. In the non-Umklapp model, the backward scattering dominates. As a result, two electrons with opposite momentum show stronger correlations. The overall correlation of the non-Umklapp model is significantly weaker than the former two models. It suggests that the Umklapp scattering dominates the low-energy correlations of the Hubbard model.

\begin{figure}[htbp]
\centering
\includegraphics[width=0.45\textwidth]{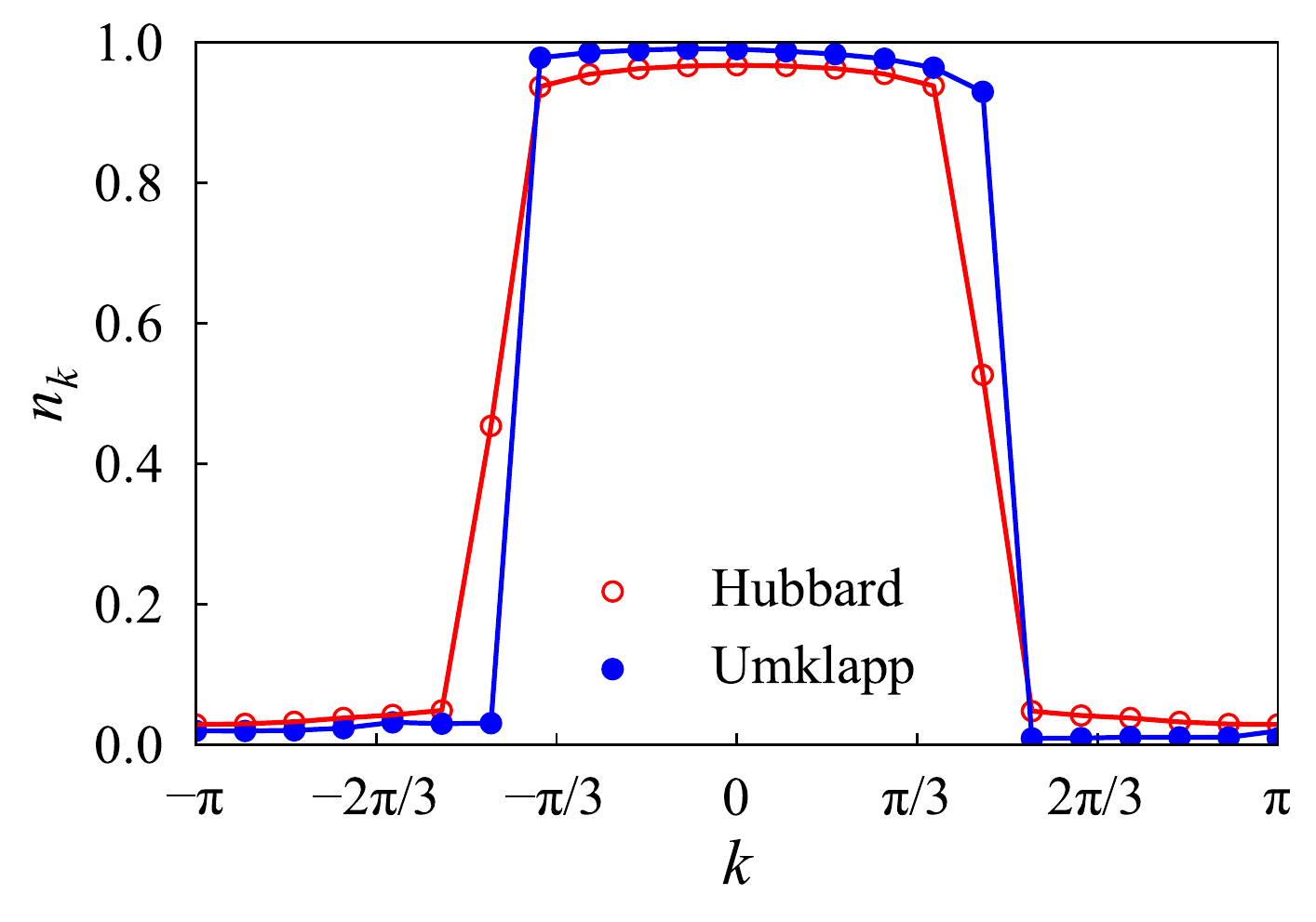}
\caption{
 Momentum distribution functions $n_k$ of electrons in the ground state of the Hubbard and Umklapp ($U_2=U$) models with $L+2$ electrons. $D=4000$ basis states are retained in the kDMRG calculation. $U=2.5t$ and $L=22$. 
 }
\label{Fig:nk}
\end{figure}

\end{document}